\documentclass[prl,twocolumn,twoside,showpacs,superscriptaddress,floatfix]{revtex4}
\usepackage{graphics,amssymb,amsmath}

\begin{document}

\title{Factors that predict better synchronizability on complex networks}
\author{H. Hong}
\affiliation{School of Physics, Korea Institute for Advanced Study, Seoul 130-722, Korea}
\author{Beom Jun Kim}
\affiliation{Department of Molecular Science and Technology, Ajou University, Suwon 442-749, Korea}
\author{M. Y. Choi}
\affiliation{Department of Physics, Seoul National University, Seoul 151-747, Korea}
\affiliation{School of Physics, Korea Institute for Advanced Study, Seoul 130-722, Korea}
\author{Hyunggyu Park}
\affiliation{School of Physics, Korea Institute for Advanced Study, Seoul 130-722, Korea}

\date{\today}

\begin{abstract}
While shorter characteristic path length has in general been believed to enhance
synchronizability of a coupled oscillator system on a complex network,
the suppressing tendency of the heterogeneity of the degree distribution,
even for shorter characteristic path length, has also been reported.
To see this, we investigate the effects of various factors such as
the degree, characteristic path length, heterogeneity, and betweenness centrality
on synchronization, and find a consistent trend between the synchronization
and the betweenness centrality. The betweenness centrality is thus proposed
as a good indicator for synchronizability.
\end{abstract}


\maketitle

In recent years, diverse systems in nature have been observed to exhibit the
characteristics of complex networks, drawing much attention to complex
network systems~\cite{ref:review,ref:network,ref:WS}.
Previous studies have mostly been focused on structural properties
of the networks rather than dynamical ones, even though dynamical properties
are also very important for understanding the systems on complex networks.
In a few studies~\cite{ref:WS,ref:synch_Hong,ref:synch_Pecora}, on the other hand,
the dynamical system of coupled oscillators has been considered
on complex networks and collective synchronization displayed by the system
has been investigated. There it has been found that shorter characteristic
path length tends to enhance synchronization.
In contrast to this, a recent paper~\cite{ref:hetero},
investigating the effects of heterogeneity of the degree distribution
on collective synchronization, reported that synchronizability is suppressed as the
degree distribution becomes more heterogeneous, even for shorter
characteristic path length.
These different results then raise a question as to synchronization on
complex networks:
What is the most important ingredient for better synchronizability?

As an attempt to give an answer to this,
we in this paper consider a system of coupled limit-cycle oscillators
on the Watts-Strogatz (WS) small-world network~\cite{ref:WS}, and
investigate collective synchronization of the system.
We pay particular attention to how the synchronization is affected by various
factors such as the maximum degree, characteristic path length,
heterogeneity of the degree distribution, and betweenness centrality.
Here the collective synchronization is explored via
the eigenvalues of the coupling matrix, which describes the
stability of the fully synchronized state~\cite{ref:synch_Pecora,ref:eigenvalue}.

The WS small-world network is constructed in the following way~\cite{ref:WS}:
We first consider a one-dimensional regular network of $N$ nodes under periodic
boundary conditions, with only local connections of range $r$ between the nodes.
Next, each local link is visited once, and with the rewiring probability $p$
it is removed and reconnected to a randomly chosen node.
At each node of the small-world network built as above, an oscillator is placed; a link
connecting two nodes represents coupling between the two oscillators at those two
nodes.
We now investigate the synchronization of the coupled oscillators on the small-world
network with given $r$ and $p$. Describing the state of the $i$th oscillator (i.e., the
one at node $i$) by $x_i$, we begin with the set of equations of motion governing
the dynamics of the $N$ coupled oscillators:
\begin{equation}
\dot x_i = F(x_i) + K\sum_{j=1}^{N} M_{ij}G(x_j),
\label{eq:linear}
\end{equation}
where $\dot x_i=F(x_i)$ governs the dynamics of individual oscillators
(i.e., with coupling strength $K=0$) and $G(x_j)$ makes the output function.
The $N\times N$ coupling matrix $M_{ij}$ is given by
\begin{equation} \label{eq:Dij}
M_{ij} = \left\{
\begin{array}{cl}
k_i & \mbox{for $i=j$} \\
-1 & \mbox{for $j\in \Lambda_i$} \\
0 & \mbox{otherwise},
\end{array}
\right.
\end{equation}
which lacks the translational symmetry due to the presence of shortcuts on the
WS small-world network.
In the case of a locally-coupled (hypercubic) network with the coordination number
$z\,(= 2D)$, the coupling matrix $M_{ij}$ has the value $z$ on the diagonal
and $-1$ on the $z$ off-diagonals adjacent to the diagonal.

The eigenvalues of the coupling matrix have been widely used to determine the linear
stability of the fully synchronized state
($x_1=x_2=\cdots =x_N$)~\cite{ref:synch_Pecora,ref:eigenvalue}.
Whereas the smallest eigenvalue, denoted by $\lambda_0$, is always
zero, the ratio of the maximum eigenvalue $\lambda_{\rm max}$ to the
smallest {\em non-vanishing}
one $\lambda_{\rm min}$ may be used as a measure for the stability
of the synchronized state, with larger values of the ratio
$\lambda_{\rm max}/\lambda_{\rm min}$
corresponding to poor
synchronizability~\cite{ref:synch_Pecora,ref:eigenvalue}.
For a general $D$-dimensional hypercubic network of linear size $L$,
the eigenvalues are given by
\begin{equation}
\lambda_{\{m_{\alpha}\}} = 4\sum_{\alpha =1}^{D} \sin^2{\frac{\pi m_{\alpha}}{L}},
\end{equation}
where $m_{\alpha} = 0,1,\cdots, L{-}1$ and $L^D \equiv N$.
It is then obvious that the eigenvalue ratio behaves as
$\lambda_{\rm max}/\lambda_{\rm min} \sim N^{2/D}$ and grows large in the thermodynamic limit
$(N\rightarrow\infty)$.
It is thus concluded that the fully synchronized state is not stable
in any $D$-dimensional regular network.

\begin{figure}
\centering{\resizebox*{!}{5.7cm}{\includegraphics{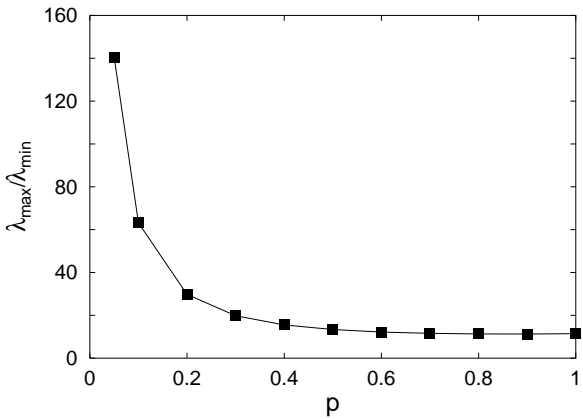}}}
\caption{Behavior of the ratio of the maximum eigenvalue $\lambda_{\rm max}$
to the smallest non-vanishing eigenvalue $\lambda_{\rm min}$ with the rewiring
probability $p$. As $p$ is raised, the ratio is shown to reduce,
yielding better synchronization.
}
\label{fig:eigen}
\end{figure}

The eigenvalue ratio for the WS network of nodes $N=2000$ and range $r=3$ is obtained numerically
and its behavior with the rewiring probability is exhibited in Fig.~\ref{fig:eigen},
where the average has been taken over 100 different network realizations.
The network size $N$ has been varied from $100$ to $2000$, only to give no
qualitative difference.
As the rewiring probability $p$ is increased, the ratio $\lambda_{\rm max}/\lambda_{\rm min}$
is observed to decrease, which implies enhancement of synchronizability.

\begin{figure}
\centering{\resizebox*{!}{5.7cm}{\includegraphics{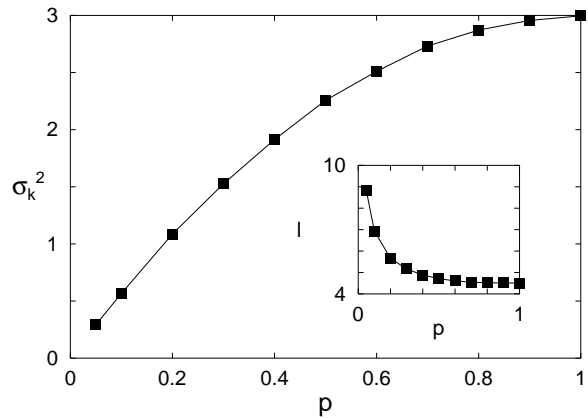}}}
\caption{Behavior of the variance ${\sigma_k}^2$ of the degree
distribution with the rewiring probability $p$.
Inset: characteristic path length $\ell$ vs. the rewiring probability $p$.
Here and in subsequent figures lines are merely guides to eyes.
}
\label{fig:dk_d}
\end{figure}

To explore how structural properties of the underlying network
affect synchronization of the system, we now examine such properties as
the characteristic path length, the betweenness centrality, and
the variance of the degree distribution, which is a measure of the heterogeneity of the
degree distribution.
Figure~\ref{fig:dk_d} displays the behavior of the variance
${\sigma_k}^2 \equiv \langle N^{-1}\sum_{i}{k_i}^2 \rangle - \langle(N^{-1}\sum_{i}k_i)^2 \rangle $
of the degree distribution for the WS small-world network with the same size $N=2000$ and range $r=3$
as that in Fig.~\ref{fig:eigen}, depending on the rewiring probability $p$.
As $p$ is increased, the variance ${\sigma_k}^2$ grows, which implies
that the degree distribution becomes more broad and heterogeneous
and that nodes of larger degrees appear.
In the inset of Fig.~\ref{fig:dk_d} the behavior of the characteristic path
length~\cite{ref:Barrat}
\begin{equation}
\ell\equiv \Bigg\langle \frac{1}{N(N-1)}\sum_{i,j}d_{i,j}\Bigg\rangle,
\end{equation}
where $\langle \cdots\rangle$ denotes the average over different realizations of the network
and $d_{i,j}$ the length of the geodesic between $i$ and $j$,
is shown as a function of the rewiring probability $p$.
As expected, the characteristic path length $\ell$ is observed to decrease
as the heterogeneity of the degree distribution (or the rewiring probability) is increased.
The results shown in Figs.~\ref{fig:eigen} and~\ref{fig:dk_d} imply that the synchronizability
on the WS network is improved as the heterogeneity of the degree distribution is increased or
as the characteristic path length is decreased, which differs from the behavior
observed in scale-free networks~\cite{ref:hetero}.

The synchronizability has been shown to be related with the load or 
betweenness centrality on nodes~\cite{ref:hetero}.
The betweenness centrality of node $n$ is defined to be~\cite{ref:Freeman,ref:Newman,ref:Holme,ref:Kahng}:
\begin{equation}
B_n \equiv \sum_{(i,j)}\frac{g_{inj}}{g_{ij}},
\end{equation}
where $g_{ij}$ is the number of geodesic paths between nodes $i$ and $j$ and
$g_{inj}$ is the number of paths between $i$ and $j$ passing through node $n$.
The summation is to be performed over all pairs of nodes $(i, j)$ such that
$i, j\neq n$ and $i\neq j$.

To get an idea of the betweenness centrality, which measures how many geodesics pass
through a given node, we first consider locally-coupled regular networks,
for which the average betweenness centrality is given by~\cite{ref:Holme}
\begin{equation}
\bar{B} \equiv \frac{1}{N}\sum_{n}B_n = (N-1)(\ell-1).
\end{equation}
Among the $N$ values of $B_n$'s, the maximum value $B^{\rm max}$
has been shown to be related with synchronizability~\cite{ref:hetero}, although
this close relation has not been stressed before.
For a $D$-dimensional local regular network, we have $\bar{B}={B}^{\rm max}$
and the characteristic path length $\ell \sim N^{1/D}$, which yields
\begin{equation}
{B}^{\rm max}\sim N^{(D+1)/D},
\end{equation}
i.e., ${B}^{\rm max}\sim N^2 , \, N^{3/2}$, and $N^{4/3}$ for
the spatial dimension $D=1, \, 2$, and $3$, respectively.
Thus the maximum value ${B}^{\rm max}$ increases algebraically with the size $N$
of the regular network, although the exponent reduces with the spatial dimension $D$.
\begin{figure}
\centering{\resizebox*{!}{5.7cm}{\includegraphics{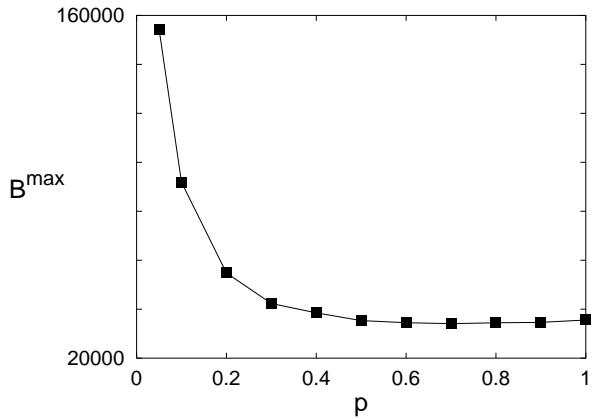}}}
\caption{Behavior of the maximum value $B^{\rm max}$ of the betweenness
centrality with the rewiring probability $p$.
${B}^{\rm max}$ is shown to decrease as $p$ is increased,
corresponding to more heterogeneous degree distributions.
}
\label{fig:b_max}
\end{figure}

\begin{figure}
\centering{\resizebox*{!}{5.7cm}{\includegraphics{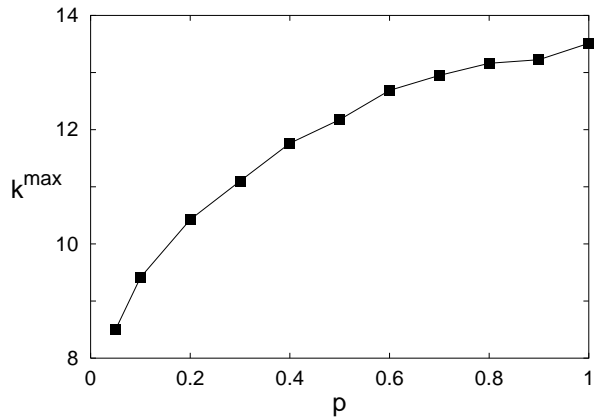}}}
\caption{The maximum degree $k^{\rm max}$ versus the rewiring probability $p$
on a WS small-world network. It is shown that $k^{\rm max}$ increases as the degree
distribution becomes more heterogeneous.
}
\label{fig:k_max}
\end{figure}

Returning to the WS small-world network, we compute the betweenness centrality $B_n$
via a modified version of the breadth-first search algorithm~\cite{ref:Newman}.
We then obtain the maximum value $B^{\rm max}$ at various values of the
rewiring probability, and display the result for a network of nodes $N=2000$ in Fig.~\ref{fig:b_max}, where the average has been taken over 100
different network realizations.
The number of nodes $N$ has also been varied from $N=100$ to $N=2000$, which does not yield
any qualitative difference.
Figure~\ref{fig:b_max} shows that the maximum load on a node reduces as
more shortcuts are introduced.
In general, on a usual scale-free network, larger values of $B^{\rm max}$
correspond to larger values of the degree.
For comparison, we also investigate the maximum degree $k^{\rm max}$
on our small-world network and display in Fig.~\ref{fig:k_max} its behavior
with the rewiring probability $p$. The increase of ${k}^{\rm max}$ with $p$
indicates the opposite trend between $B^{\rm max}$ and ${k}^{\rm max}$,
unlike the case of a scale-free network.
Note also that Fig.~\ref{fig:k_max} together with Fig.~\ref{fig:dk_d} implies
the increase of the maximum degree with the heterogeneity of the degree
distribution.

The results shown in Figs.~\ref{fig:eigen} to \ref{fig:k_max} lead to
the conclusion that synchronization on the WS network is enhanced as the heterogeneity of the
degree distribution is increased, as the characteristic path length decreased,
as the maximum betweenness centrality decreased, or as the maximum degree increased.
Remarkably, the effects of the heterogeneity of the degree distribution
as well as those of the characteristic path length (see Fig.~\ref{fig:dk_d})
differ from the results for other classes of networks~\cite{ref:hetero}.
Namely, in the case of the network studied here, larger heterogeneity of the degree 
distribution or shorter characteristic path length does not improve 
synchronizability.
On the other hand, the effects of the maximum betweenness centrality $B^{\rm max}$
appear to be consistent with those in Ref.~\cite{ref:hetero}:
Synchronizability is always improved as $B^{\rm max}$ is reduced.
Accordingly, the betweenness centrality is proposed as a suitable indicator for
predicting synchronizability on complex networks.
Regarding the maximum degree $k^{\rm max}$, Fig.~\ref{fig:k_max}
(together with Fig.~\ref{fig:eigen}) indicates that synchronizability enhances
with $k^{\rm max}$. This behavior on the WS small-world network is also in contrast with
that on a usual scale-free network, where $B^{\rm max}$ and $k^{\rm max}$ behave
similarly~\cite{ref:Kahng}, and accordingly, smaller values of $k^{\rm max}$ are
expected to give better synchronizability.

Then why does the maximum value $B^{\max}$ of the betweenness centrality strongly
affect synchronizability on networks?
An intuitive argument goes as follows: Suppose that two groups of highly linked
nodes are connected via a few nodes. Among those a few connecting nodes,
the information of the synchronized state passes through the node which has
the maximum value of the betweenness centrality, $B^{\max}$. It tends to
get overloaded since most paths go through it, which in turn leads to loss of
information. Accordingly, synchronizability is expected to become reduced as
the value $B^{\max}$ is increased. 
This argument is consistent with that of Ref.~\cite{ref:hetero}.
To check this argument, we have examined
synchronizability on the WS network before and after the removal of the node
having $B^{\max}$, by means of the eigenvalue ratio
$\lambda_{\rm max}/\lambda_{\rm min}$. In Fig. 5, the difference
$\delta \equiv (\lambda_{\rm max}/\lambda_{\rm min})_{\rm after~removal}
- (\lambda_{\rm max}/\lambda_{\rm min})_{\rm before~removal}$
is plotted against the rewiring probability $p$. Squares in Fig. 5 represent
the difference $\delta$ for the removal of the node having $B^{\max}$, and
indicate negative values regardless of the rewiring probability $p$. Those
negative values imply that $\lambda_{\rm max}/\lambda_{\rm min}$ is
reduced after the removal of the node, which in turn implies that synchronizability
of the system is enhanced after the removal of the node. In sharp contrast, 
circles in Fig. 5, which represent the difference $\delta$ for the removal of a node other than
the one with $B^{\max}$, indicate values near zero.
This implies that the eigenvalue ratio remains almost the same when an arbitrary node
is removed and accordingly that synchronizability of the system is not much affected
by the removal of an arbitrary node.
The result displayed in Fig. 5 thus manifests that the node with $B^{\max}$ is 
closely related with the synchronizability of the system whereas any other 
node is not substantially related.
\begin{figure}
\centering{\resizebox*{!}{5.7cm}{\includegraphics{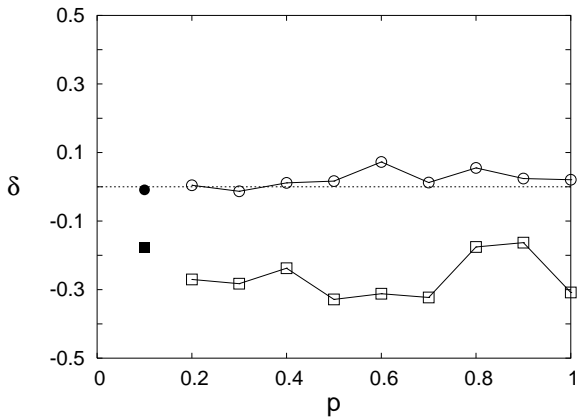}}}
\caption{
Behavior of the difference $\delta$ of the eigenvalue ratio
is displayed with the rewiring probability $p$. The data (open circles and squares)
have been obtained for the network with the number of nodes $N=2000$ and $r=3$.
Due to strong finite size effects at $p=0.1$, we use the data (filled
circle and square) for large system size of $N=5000$.
See the text for the explanation of the data symbols.
}
\label{fig:remove_Bmax}
\end{figure}

In conclusion, better synchronizability for the WS small-world network is induced
as the heterogeneity of the degree distribution is increased,
as the characteristic path length is decreased,
as the maximum betweenness centrality is reduced,
or as the maximum degree is raised.
We have found that the effects of the characteristic path length and of 
the heterogeneity on synchronization in the WS small-world network 
are different from those in the networks considered in 
Ref.~\cite{ref:hetero}. These differences seem to be related with 
the presence of hub structures in the network,  which is under further 
investigation. 
Our result implies that shorter characteristic path length or larger
heterogeneity does not always enhance the synchronizability of the
coupled system on a network. 
On the other hand, it has been observed that synchronization
is always enhanced as the betweenness centrality, measuring the load on a node,
is reduced, which is consistent with the recent result for various networks~\cite{ref:hetero}.
We have also numerically investigated the effects of the node of the maximum betweenness
centrality $B^{\max}$ on synchronizability, and found that the node of $B^{\max}$ is
highly related with the synchronizability of the system, which supports the main conclusion
of this paper.
It is thus concluded that among the important factors for better synchronization on complex
networks is a small value of the maximum betweenness centrality, rather than short
characteristic path length or large heterogeneity of the degree distribution.

This work was supported in part by the KOSEF Grant Nos. R14-2002-062-01000-0 (B.J.K.)
and R01-2002-000-00285-0 as well as by the Ministry of Education through the BK21 Project (M.Y.C.).
Numerical works were mainly performed on the computer cluster Iceberg at Ajou University.

\end{document}